\documentclass[aps,prl,twocolumn,superscriptaddress,bibliography]{revtex4-2}

\usepackage{amsfonts}
\usepackage{mathrsfs}
\usepackage{amsmath}
\usepackage{color}
\usepackage{graphicx}
\usepackage{bm}
\usepackage{amssymb}
\usepackage{xspace}
\usepackage{epstopdf}
\usepackage{dcolumn}
\usepackage{longtable}
\usepackage{multirow}
\usepackage{float}
\usepackage{comment}

\DeclareUnicodeCharacter{2212}{-}   

\usepackage[colorlinks=true, letterpaper=true, pdfstartview=FitV,  linkcolor=blue, citecolor=blue, urlcolor=blue]{hyperref}

\begin{document}

\title{In-plane anomalous Hall effect in $\mathcal{PT}$-symmetric antiferromagnetic materials}

\author{Jin Cao}
\thanks{These authors contributed equally to this work}
\affiliation{Centre for Quantum Physics, Key Laboratory of Advanced Optoelectronic Quantum Architecture and Measurement (MOE), School of Physics, Beijing Institute of Technology, Beijing, 100081, China }
\affiliation{Beijing Key Lab of Nanophotonics \& Ultrafine Optoelectronic Systems, School of Physics, Beijing Institute of Technology, Beijing, 100081, China }
\author{Wei Jiang}
\thanks{These authors contributed equally to this work}
\affiliation{Centre for Quantum Physics, Key Laboratory of Advanced Optoelectronic Quantum Architecture and Measurement (MOE), School of Physics, Beijing Institute of Technology, Beijing, 100081, China }
\affiliation{Beijing Key Lab of Nanophotonics \& Ultrafine Optoelectronic Systems, School of Physics, Beijing Institute of Technology, Beijing, 100081, China }
\author{Xiao-Ping Li}
\affiliation{Centre for Quantum Physics, Key Laboratory of Advanced Optoelectronic Quantum Architecture and Measurement (MOE), School of Physics, Beijing Institute of Technology, Beijing, 100081, China }
\affiliation{Beijing Key Lab of Nanophotonics \& Ultrafine Optoelectronic Systems, School of Physics, Beijing Institute of Technology, Beijing, 100081, China }
\author{Daifeng Tu}
\affiliation{Anhui Key Laboratory of Condensed Matter Physics at Extreme Conditions, High Magnetic Field Laboratory, HFIPS, Anhui, Chinese Academy of Sciences, Hefei 230031, P.R. China}
\affiliation{Department of Physics, University of Science and Technology of China, Hefei 230026, P.R. China }
\author{Jiadong Zhou}
\affiliation{Centre for Quantum Physics, Key Laboratory of Advanced Optoelectronic Quantum Architecture and Measurement (MOE), School of Physics, Beijing Institute of Technology, Beijing, 100081, China }
\affiliation{Beijing Key Lab of Nanophotonics \& Ultrafine Optoelectronic Systems, School of Physics, Beijing Institute of Technology, Beijing, 100081, China }
\author{Jianhui Zhou}
\email{jhzhou@hmfl.ac.cn}
\affiliation{Anhui Key Laboratory of Condensed Matter Physics at Extreme Conditions, High Magnetic Field Laboratory, HFIPS, Anhui, Chinese Academy of Sciences, Hefei 230031, P.R. China}
\author{Yugui Yao}
\email{ygyao@bit.edu.cn}
\affiliation{Centre for Quantum Physics, Key Laboratory of Advanced Optoelectronic Quantum Architecture and Measurement (MOE), School of Physics, Beijing Institute of Technology, Beijing, 100081, China }
\affiliation{Beijing Key Lab of Nanophotonics \& Ultrafine Optoelectronic Systems, School of Physics, Beijing Institute of Technology, Beijing, 100081, China }


\begin{abstract}

Anomalous Hall effect (AHE), a protocol of various low-power dissipation quantum phenomena and a fundamental precursor of intriguing topological phases of matter, is usually observed in $\textit{ferromagnetic}$ materials with orthogonal configuration between the electric field,  magnetization and the Hall current. Here, based on the symmetry analysis, we find an unconventional AHE induced by the in-plane magnetic field (IPAHE) via spin-canting effect in $\mathcal{PT}$ symmetric antiferromagnetic (AFM) systems, featuring a linear dependence of magnetic field and 2$\pi$ angle periodicity with a comparable magnitude as conventional AHE. 
We demonstrate the key findings in the known AFM Dirac semimetal CuMnAs and a new kind of AFM heterodimensional VS$_2$-VS superlattice with a nodal-line Fermi surface and also briefly discuss the experimental detection.
Our work provides an efficient pathway to search and/or design realistic materials for novel IPAHE that could greatly facilitate their application in AFM spintronic devices.
\end{abstract}

\maketitle

{\emph{\textcolor{blue}{Introduction.}}}
The anomalous Hall effect (AHE) has played a vital role in understanding low-power dissipation quantum phenomena associated with Berry phase~\citep{rmp_xiaodi_berry,Sinova2015RMP,Baltz2018RMP} and  the development of the field of topological phases of matter~\citep{Hasan2010RMP,Weng2015AP,Bansil2016RMP,Armitage2018RMP,Lv2021RMP}. 
In analogy to the conventional Hall effect \citep{Hall1897}, where the magnetic field, electric field, and Hall current are orthogonal to each other due to the Lorentz force, the AHE was commonly considered in the same manner by replacing the magnetic field by static magnetization in ferromagnetic (FM) materials \citep{Karplus1954}. 
However, the essential ingredient for AHE, i.e., time-reversal ($\mathcal{T}$) symmetry breaking, does not enforce any constraint on the magnetization direction or related magnetisms~\citep{rmp_nagaosa_ahc,Liu2016}. It in principle allows an unconventional AHE with the Hall current flows in the plane spanned by the magnetization/magnetic field and the electric field, that is, in-plane AHE (IPAHE). Due to its unique in-plane configurations, IPAHE could potentially revolutionize promising spintronic applications similar to the unidirectional spin Hall magnetoresistance \citep{Avci2015,nc2018_Yang}. 

Several theoretical investigations based on the ideal two-dimensional (2D) electron gas and a few artificially designed 2D material systems have shown the possible quantized version of IPAHE with the broken of both $\mathcal{T}$ and mirror ($\mathcal{M}$) symmetries~\citep{prl2013_Liu,prb2017_Zhong,sun2022quantized,prl2018_Liu}, most of which focus on FM or ferrimagnetic materials with an intrinsic in-plane magnetization.
Meanwhile, recent theoretical and experimental studies have reached a consensus on the Berry curvature contribution to the intrinsic AHE due to spin-orbit coupling (SOC)~\citep{prl2018_Liu,Ajaya2016,nakatsuji2015}, leading to the discovery of the AHE in antiferromagnetic (AFM) systems with a zero net magnetization~\citep{Chen2014,Martin2008,Ajaya2016,nakatsuji2015,Shao2020PRL}. It greatly inspires us to study IPAHE among the large family of AFM systems, such as the $\mathcal{PT}$ symmetric AFM materials, usually with only nonzero high-order AHE but vanishing first-order AHE~\citep{prl2021_liuhuiying_2nd,wang2021_nlHE_AFM,Smejkal2022NRM,Hayami2022_PRB_nlSHE_PTAFM}. In reality, both the essential criterion to realize IPAHE and generally applicable searching and designing strategies for promising IPAHE materials are still lacking.

In this Letter, we find a universal symmetry rule to achieve IPAHE, i.e., the absence of both rotation and reflection symmetries in at least two directions for the corresponding magnetic point group (MPG). We further elaborate on our key findings in one collinear ($\mathcal{PT}$ symmetric) AFM model system with only one direction having mirror reflection symmetry, which shows a noticeable AHE induced by an in-plane magnetic field. The complete list of MPGs allowing IPAHE is given. To facilitate material discovery, we provide two feasible approaches, i.e., the top-down and bottom-up approaches, to design realistic AFM materials that support IPAHE. In addition, the experimental detection and verification of IPAHE with two representative examples are discussed.

\begin{figure}
\begin{centering}
\includegraphics[width=8.4cm]{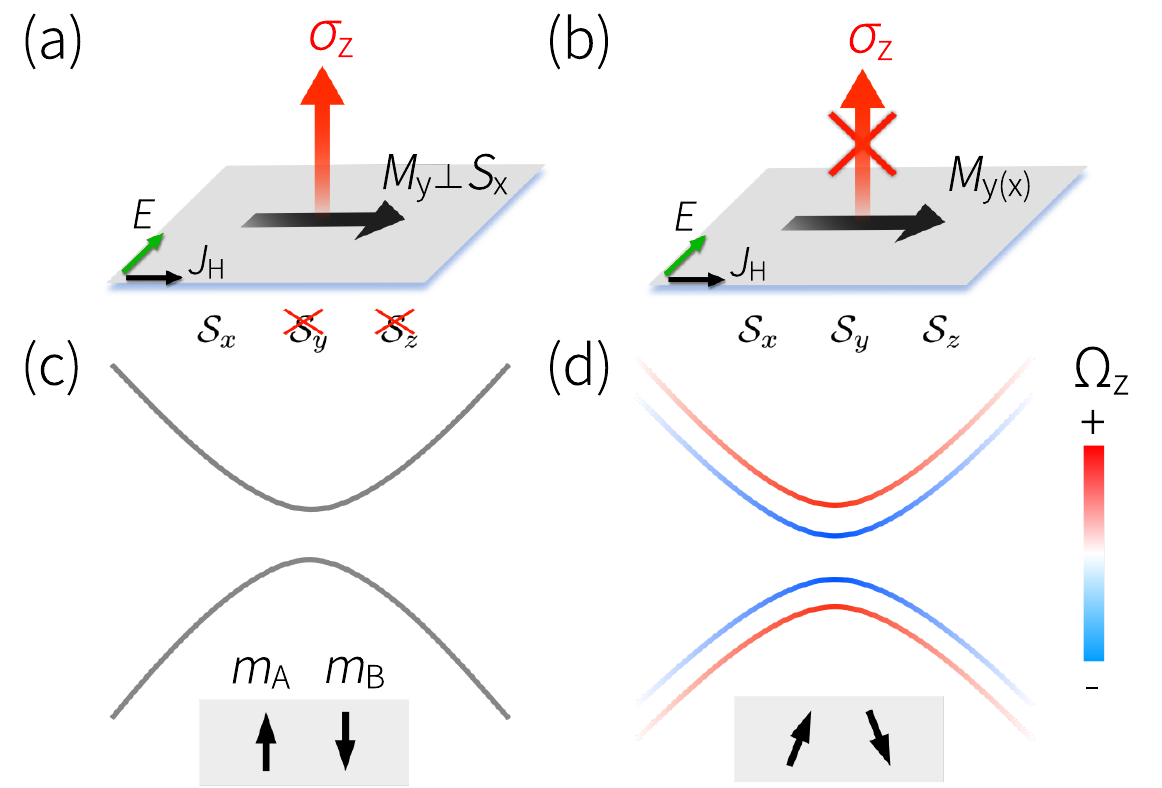}
\par\end{centering}
\caption{\label{Fig_1}Schematic rule for generating the in-plane AHE (IPAHE) in $\mathcal{PT}$ symmetric AFM systems. (a) The IPAHE $\boldsymbol{J}_{H}=\boldsymbol{E}\times\boldsymbol{\sigma}$ is generated when the orthogonal $\sigma_{z}$ and $M_{y}$ are perpendicular to the direction of $\mathcal{S}_{x}$ (i.e., $x$ direction). Here, $\mathcal{S}_{x}$ represents mirror $\mathcal{M}_{x}$, rotation $\mathcal{C}_{nx}$, and their combination with $\mathcal{T}$ ($\mathcal{TM}_{x}$ and $\mathcal{TC}_{nx}$). (b) Any two directions of $\mathcal{S}_{\gamma} (\gamma=x,y,z)$ together enforce zero IPAHE. (c) The doubly degenerated band structure in the absence of an external magnetic field shows vanishing Berry curvature. The collinear AFM state is shown in black arrows. (d) The net magnetization is induced through the spin-canting effect. It lifts the band degeneracy, showing the opposite Berry curvature in the small-gap region, giving rise to the IPAHE.  
}
\end{figure}

{\emph{\textcolor{blue}{Symmetry analysis.}}}
The intrinsic anomalous Hall conductivity (AHC) evaluated by the integration of the Berry curvature over the Brillouin zone, is usually proportional to the magnetization~\citep{Yao2004PRL,ZengCG2006PRL}. By expanding AHC up to the linear order of magnetization, one can define the rank-two tensor
\begin{equation}
    \chi_{\alpha \beta}=\frac{\partial \sigma_{\alpha}}{\partial M_{\beta}},
    \label{Eq_pheo}
\end{equation}
where $M_{\beta}$ denotes the magnetic moment per unit cell, and $\sigma_{\alpha} = \frac{1}{2}\epsilon_{\alpha\beta\gamma} \sigma_{\beta\gamma}$ is the Hall pseudovector, with the Levi-Civita symbol $\epsilon_{\alpha\beta\gamma}$ defining the antisymmetric dissipationless parts of the linear conductivity tensor $\sigma_{\beta\gamma}$. 
The tensor $\chi$ by definition is independent of $M_{\beta}$, and should respect the symmetry of the system at $M_{\beta}=0$. The diagonal part of $\chi$ gives the normal out-of-plane AHC, e.g., $\chi_{zz}$~\citep{LandauEDC}, whereas the off-diagonal parts, e.g., $\chi_{zy}$, represent the IPAHE~\citep{pheahe}. Therefore, the tensor $\chi$ provides a complete description of the intrinsic AHE in magnetic materials.

The normal AHE cannot be ruled out by any structural symmetry operations, while the IPAHE requires low symmetry of the system \citep{SM}. Consider a class of MPG operations, $\mathcal{S}_{\gamma}$, which represents either mirror $\mathcal{M}_{\gamma}$, rotation $\mathcal{C}_{n\gamma}$, or their combination with $\mathcal{T}$ ($\mathcal{TM}_{\gamma}$ and $\mathcal{TC}_{n\gamma}$), where $\gamma$ is the direction of the corresponding symmetry operations. Given that both $\sigma_{\alpha}$ and $M_{\beta}$ are pseudovectors that transform in the same way under $\mathcal{S}_{\gamma}$, they should be simultaneously parallel or perpendicular to the direction of $\mathcal{S}_{\gamma}$ to avoid the elimination of $\chi_{\alpha \beta}$ by $\mathcal{S}_{\gamma}$. Since we are concerned about the IPAHE, i.e., $\alpha \neq \beta$, the orthogonal $\sigma_{\alpha}$ and $M_{\beta}$ should also be perpendicular to the direction of $\mathcal{S}_{\gamma}$ to host IPAHE, as sketched in Fig.~\ref{Fig_1}(a). 
If there exists more than one direction of $\mathcal{S}_{\gamma_{i}}$, one cannot find a configuration that support nonzero $\chi_{\alpha \beta}$, where the orthogonal $\sigma_{\alpha}$ and $M_{\beta}$ is perpendicular to all of the $\gamma_{i}$ directions. 
The detailed constraint on the IPAHE tensor components under MPG operations are summarized in Table.~\ref{tab:I}.
So far, we can obtain the essential symmetry rule for IPAHE: the absence of $\mathcal{S}_{\gamma}$ in at least two directions for the corresponding MPG.

\begin{table}
\caption{\label{tab:I}Constraint on the IPAHE tensor components $\chi_{\alpha \beta}$
($\alpha \protect\neq \beta$) under $\mathcal{S}_{\gamma}$ ($\gamma=x,y,z$) operations. The mark $\checkmark$ ($\times$) denotes symmetry allowed (forbidden) components.}
\begin{ruledtabular}
\begin{tabular}{cccc}
 & $\mathcal{S}_{x}$ & $\mathcal{S}_{y}$ & $\mathcal{S}_{z}$ \\
\hline 
$\chi_{xy}$, $\chi_{yx}$ & $\times$ & $\times$ & $\checkmark$ \\
$\chi_{xz}$, $\chi_{zx}$ & $\times$ & $\checkmark$ & $\times$ \\
$\chi_{yz}$, $\chi_{zy}$ & $\checkmark$ & $\times$ & $\times$ \\
\end{tabular}
\end{ruledtabular}
\end{table}

{\emph{\textcolor{blue}{IPAHE in $\mathcal{PT}$ symmetric AFM system.}}}
The IPAHE has been reported mostly in FM or ferrimagnetic systems with a nonzero net magnetization~\citep{prl2013_Liu,prl2018_Liu,Tan2021}. Here, we focus on the $\mathcal{PT}$ symmetric AFM system with vanishing magnetization to realize IPAHE. Consider the simplest collinear AFM system, in the absence of an external magnetic field, as sketched in Fig.~\ref{Fig_1}(c), there are two magnetic sublattices denoted as $\boldsymbol{m}_{\tau}$ ($\tau=A,B$), satisfying $\boldsymbol{m}_{A}=-\boldsymbol{m}_{B}$. Due to the $\mathcal{PT}$ symmetry, the Kramer degeneracy exists in the entire Brillouin zone (BZ), leading to the vanishing AHE. Considering that the Berry curvature is usually non-Abelian, one could obtains a finite AHE by removing the Kramer degeneracy~\citep{rmp_xiaodi_berry}.

By applying an external magnetic field $\boldsymbol{B}$ perpendicular to $\boldsymbol{m}_{\tau}$, the local magnetic moments will be slightly titled accordingly due to the spin-canting effect, yielding $\boldsymbol{m}_{\tau}=\boldsymbol{m}^{(0)}_{\tau}+\boldsymbol{m}^{\prime}_{\tau}$. The net magnetic moment per unit cell will be $\boldsymbol{M}=\boldsymbol{m}_{A}^{\prime}+\boldsymbol{m}_{B}^{\prime}$, which will lift the Kramer degeneracy, and bring in a finite Berry curvature [Fig.~\ref{Fig_1}(d)].  
Note that we mainly focus on the weak magnetic field region that does not need to consider spin-flop/flip transition and the spin-canting effect can be understood as one small perturbation~\citep{Blundell}. 
Therefore, the corresponding net magnetic moment is approximately linear in the magnetic field $\boldsymbol{M}\propto\boldsymbol{B}$. 
Following the aforementioned symmetry analysis of IPAHE, it is not forbidden to achieve IPAHE in $\mathcal{PT}$ symmetric AFM systems, as also confirmed in the toy model~\citep{SM}. 

Though the $\mathcal{PT}$ symmetric AFM systems widely exist in nature~\citep{MAGNDATA_I,MAGNDATA_II}, the IPAHE in such systems has not yet been reported. To assist materials discovery,
we provide a complete list of MPGs with $\mathcal{PT}$ symmetry that allow IPAHE~\citep{SM}.
Further, to facilitate the experimental verification, we propose two general approaches to efficiently target suitable AFM materials that host IPAHE, i.e., the top-down approach to engineer existing AFM materials and the bottom-up approach to design new materials.

\begin{figure}
\begin{centering}
\includegraphics[width=8.4cm]{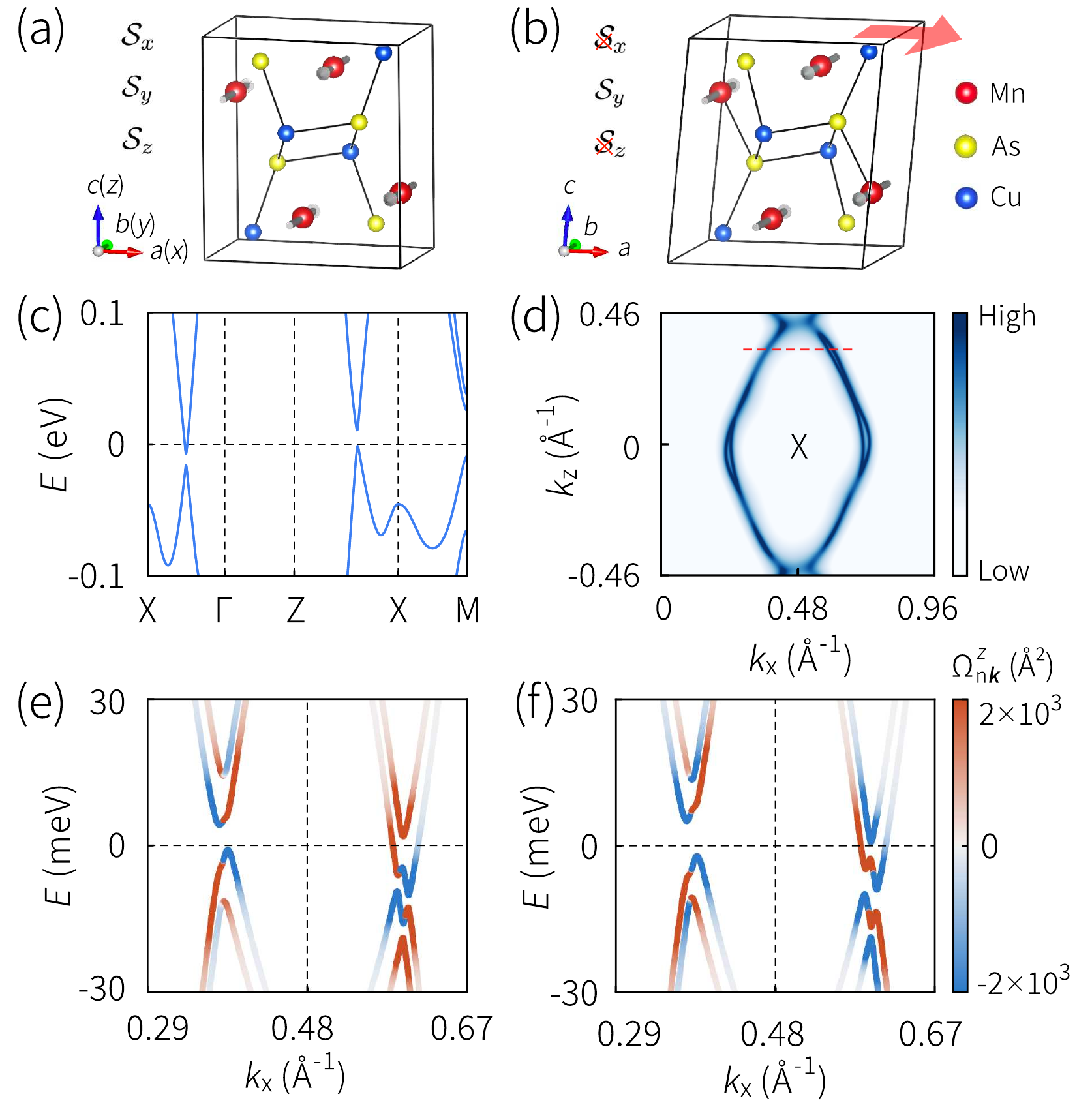}
\par\end{centering}
\caption{\label{Fig_2}The top-down approach to engineer IPAHE materials. (a) The AFM-$b$ state of orthorhombic CuMnAs has $\mathcal{S}_{\gamma}$ in three directions that prohibits the IPAHE. (b) Schematic of the CuMnAs under shear strain along the $a$ direction, which has only $\mathcal{S}_{y}$ symmetry thus allowing IPAHE. (c) The band structure after shear strain. (d) The Fermi surface plot  in the $k_y=0$ plane at $E_F=0$. (e) and (f) The Berry curvature distributions along the red dash line in panel (d) with spin-canting effect $M_x=0.4~\mu_B$ and $M_z=0.4~\mu_B$ per unit cell, respectively. }
\end{figure}

{\emph{\textcolor{blue}{Strained CuMnAs.}}}
We first demonstrate the IPAHE in well-established $\mathcal{PT}$ symmetric AFM systems. However, since many of those materials may not meet the symmetry requirement for IPAHE, it is necessary to apply feasible external field control to lower the structural symmetries to host IPAHE. Hereafter, we term this searching method the top-down approach. With the advancement of modern technologies in fine tuning of material properties, various field control methods have been successfully applied in multiple material systems, such as electric field, strains, high pressure and chemical doping \citep{Jiang2020,Jiang2021,Zhang2022}. 
Here, we would examine IPAHE in one well-known and typical $\mathcal{PT}$ symmetric AFM systems, orthorhombic CuMnAs \citep{CuMnAs_exp_1992,CuMnAs_exp_2012_room,Tang2016,Emmanouilidou2017,ifmmodeSelseSfimejkal2017,Huyen2021}, which notably exhibits significant potential in electrical switching of the AFM order \citep{Wadley2016Sicence} and terahertz electrical writing for memory \citep{Olejnik2018SciAdv}. 

The crystal structure of orthorhombic CuMnAs is shown in Fig.~\ref{Fig_2}(a), where the AFM aligned magnetic moment is mainly concentrated on the Mn atoms along the $b$ direction (AFM-$b$), as indicated by grey arrows, which has been observed in experiments~\citep{Emmanouilidou2017}. The AFM-$b$ state belongs to the MPG $m^{\prime}mm$~\citep{book_spg}, generated by $\mathcal{TM}_{x}$, $\mathcal{M}_{y}$, and $\mathcal{M}_{z}$. 
It consists of $\mathcal{S}_{\gamma}$ operations in all three directions that prohibit IPAHE. To lower the crystal symmetry, we applied a $2\%$ shear strain in the $a$ direction, as sketched in Fig.~\ref{Fig_2}(b). The new AFM-$b$ state has a lower symmetry $2^{\prime}/m$, consisting of only $\mathcal{PT}$, $\mathcal{T}\mathcal{C}_{2y}$, and $\mathcal{M}_{y}$, i.e., only $\mathcal{S}_y$. The corresponding band structure under $2\%$ shear strain is shown in Fig.~\ref{Fig_2}(c), which preserves well the nodal-line structure on the $k_y=0$ plane surrounding the X point near the Fermi level, as also confirmed by the Fermi surface plot in Fig.~\ref{Fig_2}(d). 

Due to the $\mathcal{S}_y$ symmetry, the nonzero IPAHE components are $\chi_{zx(xz)}$. The distributions of calculated Berry curvature with spin-canting effect~\citep{SM} are shown in Figs.~\ref{Fig_2}(e-f). It is clear that the splitting of the doubly degenerate bands generates a nonzero Berry curvature near the gap opening area, leading to a finite IPAHE that will be discussed in details later. 
We note that the magnetic ground state may change from AFM-$b$ to AFM-$c$ with certain amount of hole doping~\citep{Tang2016}. Thus, we also tested AFM-$c$ state and confirmed that IPAHE can also be achieved through proper strain engineering~\citep{SM}. 

\begin{figure}
\begin{centering}
\includegraphics[width=8.4cm]{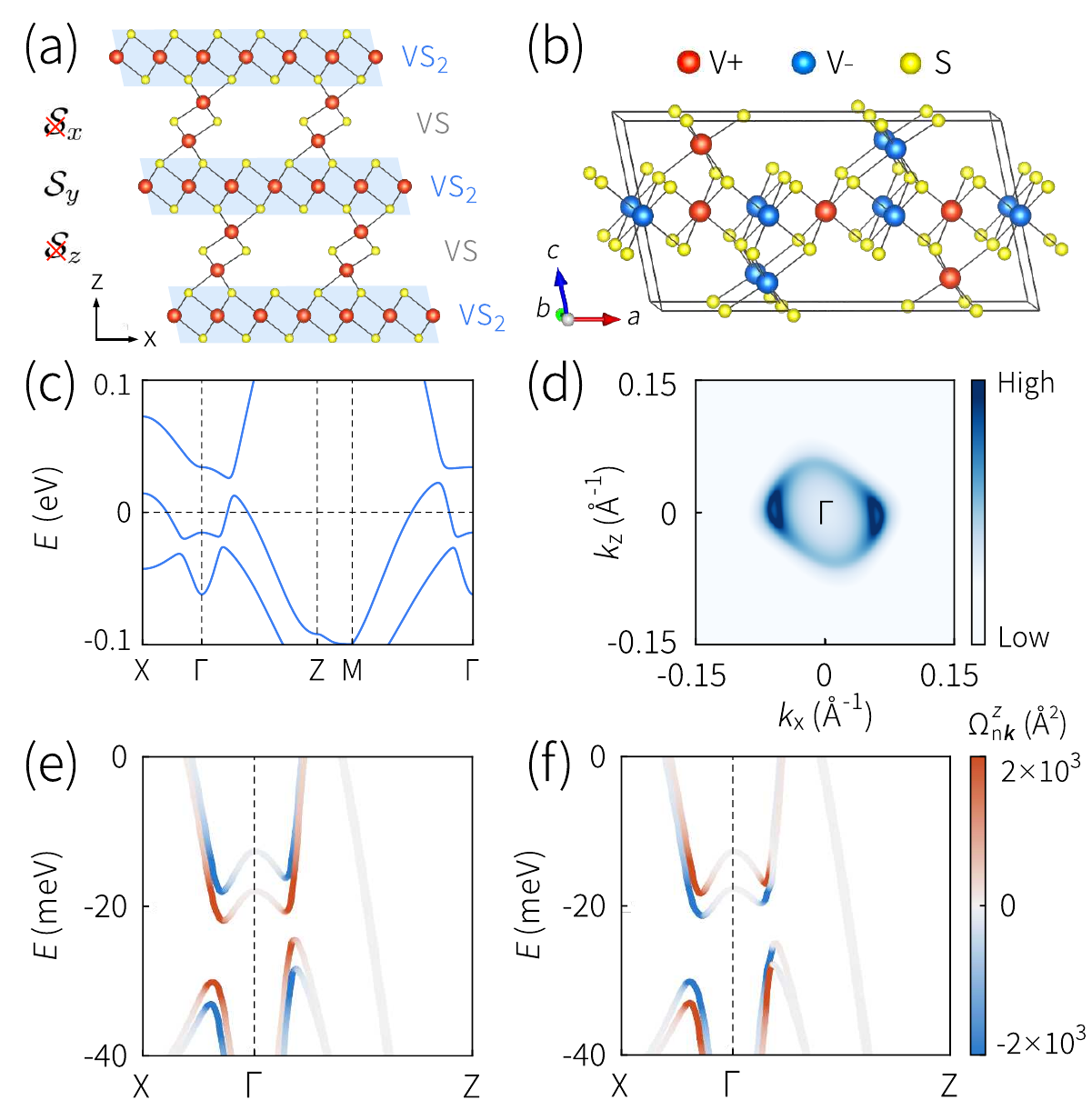}
\par\end{centering}
\caption{\label{Fig_3}The bottom-up approach to design IPAHE materials. (a) The heterodimensional superlattice consists of 2D $\mathrm{VS_2}$ and 1D VS array that naturally meets the symmetry requirement of IPAHE. (b) Sketch of the AFM-$b$ state in the VS$_2$-VS superlattice. The notation $\mathrm{V}+$ and $\mathrm{V}-$ denotes opposite local magnetic moments on the V atoms along $b$ direction. (c) The band structure and (d) The density of states in the $k_y=0$ plane at $E_F=-20$~meV. (e) and (f) The Berry curvature distributions with spin-canting effect $M_x=0.4~\mu_B$ and $M_z=0.4~\mu_B$ per unit cell, respectively. }
\end{figure}

{\emph{\textcolor{blue}{Heterodimensional VS$_2$-VS superlattice.}}}
Besides the top-down approach, one can take advantage of the recent rapid advancement of modern material synthesis through bottom-up design to create IPAHE materials, for example, in heterostructures~\citep{Zhou2022_nm,zhou_VS2VS_superlattice}. Given the relatively low symmetry of one-dimensional (1D) materials, heterostructure systems formed by a combination of 1D and 2D/3D materials could easily satisfy the aforementioned symmetry requirements. In fact, these heterostructure, i.e., heterodimensional, materials have been successfully synthesized in families of 3D transition metal chalcogenides (TMCs), where 2D transition metal dichalcogenide (TMD) layers are connected through intercalated layers constructed by the same transition metal element with a typical 1D array pattern~\citep{Zhou2022_nm,zhou_VS2VS_superlattice}. The intercalation of quasi-1D arrays could effectively reduce the symmetry of the heterodimensional system, which also provides one tunable pathway to modify the magnetization easy axis between the out-of-plane and in-plane directions. Here, we elaborate on the heterodimensional superlattice formed by a combination of 2D TMD and 1D transition-metal monochalcogenide array through bottom-up approach, as shown in Fig.~\ref{Fig_3}, and illustrate the corresponding IPAHE due to its designed structure with desired symmetry. 

To reach the required symmetry, we follow the general stacking pattern of experiments with an alternating 2D TMD and 1D array layer by layer, as shown in Fig.~\ref{Fig_3}(a). The nonmagnetic structure belongs to the $C_{2h}$ point group with only one direction of mirror symmetry $\mathcal{M}_{y}$, which perfectly meet the symmetry requirement of IPAHE. Furthermore, we selected the vanadium based materials and carried out DFT calculations~\citep{SM}. 
We found the VS$_2$-VS superlattice prefers the AFM state with the magnetization aligned along the $b$ direction~\citep{SM} as shown in Fig.~\ref{Fig_3}(b), denoted as AFM-$b$. It belongs to MPG $2/m1^\prime$ generated by $\mathcal{T}$, $\mathcal{P}$, and $\mathcal{M}_y$, i.e., only $\mathcal{S}_y$. Thus, we will use VS$_2$-VS as one representative example to demonstrate the intriguing IPAHE in heterodimensional materials.

In the absence of SOC, the band structure near the Fermi level is dominated by the nodal line on the $k_y=0$ plane. At the $\Gamma$ point, the two bands have opposite eigenvalues of $\mathcal{M}_{y}$, indicating this nodal line is protected by the $\mathcal{M}_{y}$ symmetry. After including the SOC effect, there is a small gap near the nodal line, as shown in Figs.~\ref{Fig_3}(c-d). Since the SOC effect respects the $\mathcal{PT}$, the energy bands remain doubly degenerate and the Berry curvature vanishes in the entire Brillouin zone, resulting in a vanishing AHE.

Due to the $\mathcal{S}_y$ symmetry, the nonzero IPAHE components is $\chi_{zx(xz)}$. Under the influence of spin-canting effect due to the external magnetic field, the energy bands split and the Berry curvature becomes nonzero with pronounced contribution near the gap opening area, as shown in Figs.~\ref{Fig_3}(e-f). The resulting non-zero Berry curvature leads to a finite IPAHE similar to the strained CuMnAs. The band-resolved Berry curvature distribution also implies the energy dependence of the AHC that can be potentially accessed by the electric gating or chemical doping.    

\begin{figure}
\begin{centering}
\includegraphics[width=8.4cm]{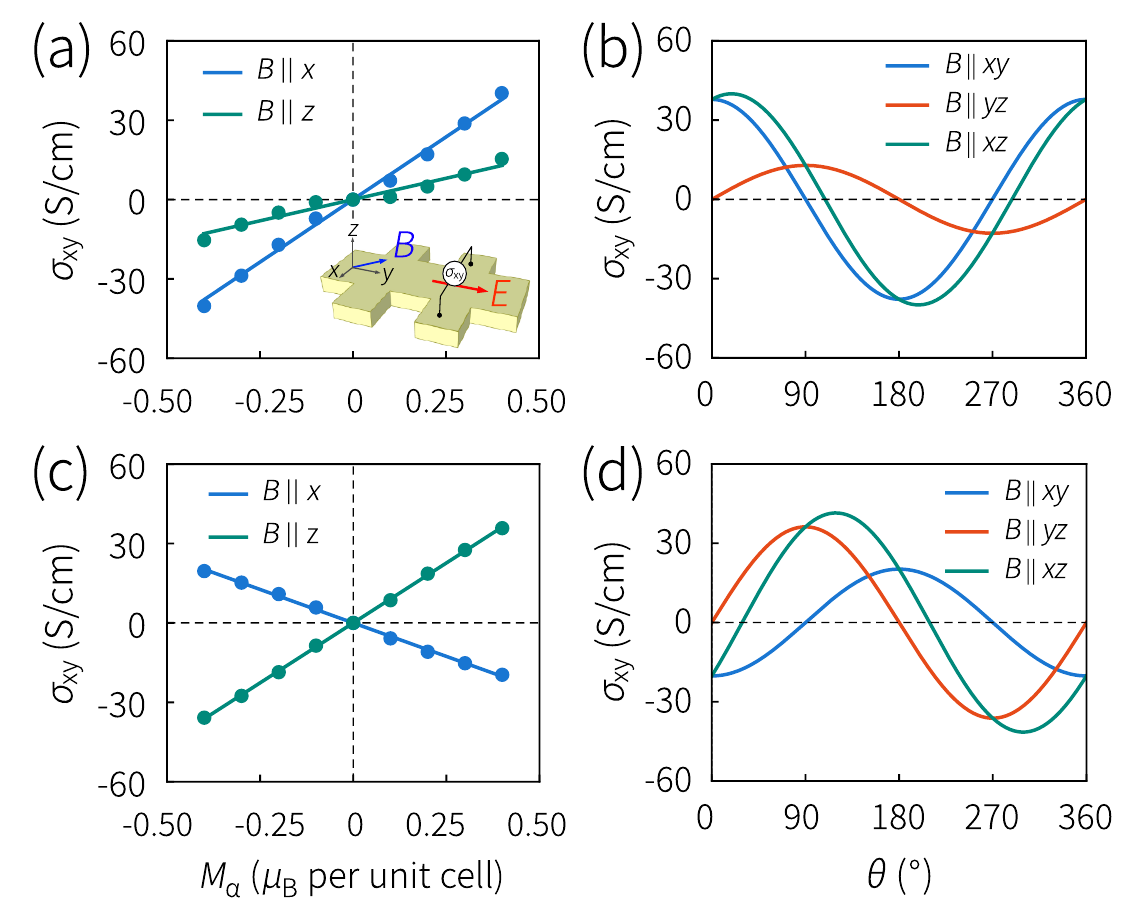}
\par\end{centering}
\caption{\label{Fig_4}The in-plane and normal AHC for strained CuMnAs in panels (a-b) and VS$_2$-VS superlattice in panels (c-d). In panels (a) and (c), the calculated results are shown as scatter diagrams, and the fitted results are given in solid lines. The panels (b) and (d) are the AHC under the arbitrary direction of the magnetic field. The Fermi levels were chosen near the nodal line. $E_F=-24$~meV and $E_F=-20$~meV were used for CuMnAs and VS$_2$-VS superlattice, respectively. }
\end{figure}

{\emph{\textcolor{blue}{Experimental signature.}}}
We turn to the experimental detection of IPAHE in a standard four-terminal Hall setup [inset of Fig.~\ref{Fig_4}(a)]. In order to observe the IPAHE while suppressing the contribution from the orbital effect due to the Lorentz force~\citep{ZimanEP}, it is desirable to choose nanoflakes of the thickness in the range from several nm to tens of nm. This can now be easily achieved experimentally through either mechanical exfoliation or chemical vapor depositions.  
The strained CuMnAs and VS$_2$-VS superlattice have only the $\mathcal{S}_y$ symmetry, which imposes the same constraint on the IPAHE. The nonzero tensor elements are $\chi_{zx}$ and $\chi_{zz}$. Then, the anomalous Hall conductivity can be given by 
\begin{eqnarray}
\sigma_{xy} & = & \chi_{zx}M_{x}+\chi_{zz}M_{z}.\label{eq:2}
\end{eqnarray}
The calculated AHC for CuMnAs and VS$_2$-VS are depicted in Figs.~\ref{Fig_4}(a) and (c), respectively, showing a linear dependence relation in $\boldsymbol{M}$, similar to the conventional AHE in ferromagnets~\citep{ZengCG2006PRL}. By fitting Eq.~(\ref{eq:2}), we obtain $\chi_{zx}=95\,\mathrm{S/\left(cm\cdot\mu_{B}\right)}$ and $\chi_{zz}=32\,\mathrm{S/\left(cm\cdot\mu_{B}\right)}$ for strained CuMnAs, and $\chi_{zx}=-51\,\mathrm{S/\left(cm\cdot\mu_{B}\right)}$ and $\chi_{zz}=91\,\mathrm{S/\left(cm\cdot\mu_{B}\right)}$ for VS$_2$-VS. Based on the fitted tensor $\chi$, the AHC under any arbitrary direction of the magnetic field can be obtained from Eq.~(\ref{eq:2}), as shown in Figs.~\ref{Fig_4}(b) and (d) for CuMnAs and VS$_2$-VS, respectively. 
In general, there may exist higher order corrections of $\boldsymbol{M}$ in Eq.~(\ref{eq:2})~\citep{hoM}.
Since the spin-canting induced magnetic moment per unit cell is very small ($M\ll 1~\mu_{B}$), such corrections could be negligible. 

It should be noted that considering the diversity of TMD materials and their combination with quasi-1D intercalated layers, such IPAHE could be widely exist in other similar materials with the same structural symmetry. We also point out that the extrinsic AHE originating from the side jump process usually shares the same scaling relation against the longitudinal conductivity as the intrinsic AHE~\citep{Berger1970PRB}. The distinct dependence of AHE on the carrier density, temperature, and types of disorders could help us to identify the the intrinsic IPAHE due to the Berry curvature in experiments~\citep{rmp_nagaosa_ahc}.

{\emph{\textcolor{blue}{Summary and discussion.}}}
We have first established a comprehensive rule of the IPAHE with minimum symmetry requirement and extend its material realization to a large family of $\mathcal{PT}$ symmetric AFM systems that have been rarely studied before \citep{MAGNDATA_I,MAGNDATA_II}. 
In fact, the mechanism for in-plane field induced AHE can be straightforwardly applicable to other Berry phase related quantum phenomena, such as the anomalous Nernst effect and anomalous thermal Hall effect. 
Meanwhile, these $\mathcal{PT}$ symmetric AFM systems also provide a promising platform to observe novel quantum phenomena and to built energy-efficient AFM spintronic devices, which should draw immediate experimental attention.

\begin{acknowledgments}
This work was supported by the National Key R\&D Program of China (Grant No.~2020YFA0308800), the NSF of China (Grants Nos.~12061131002, 11734003, 62174013), the Strategic Priority Research Program of Chinese Academy of Sciences (Grant No.~XDB30000000).
W. J. was supported by the Beijing Institute of Technology Research Fund Program for Young Scholars. 
J. Z. was supported by the High Magnetic Field Laboratory of Anhui Province.
\end{acknowledgments}


\bibliography{ref}

\begin{thebibliography}{54}
\bibitem [{hoM()}]{hoM}%
  \BibitemOpen
  \href@noop {} {\bibinfo  {journal} {Based on the Onsager's relation, we have
  $\sigma_{xy}(-\boldsymbol{M})=-\sigma_{xy}(\boldsymbol{M})$~\citep{rmp_nagaosa_ahc,SM},
  allows the higher order contribution of $M$. Thus, the leading order
  correction is of $M^{3}$}\ }\BibitemShut {NoStop}%
\end{thebibliography}

\end{document}